# Four-terminal perovskite/silicon tandem solar cell with integrated Mie-resonant spectral splitter metagrating

Verena Neder[1,2], Dong Zhang[3,4], Sjoerd Veenstra[3], and Albert Polman[2]

[1]Institute of Physics, University of Amsterdam
Science Park 904, 1098 XH Amsterdam, the Netherlands

[2]Center for Nanophotonics, NWO-Institute AMOLF
Science Park 104, 1098 XG, Amsterdam, the Netherlands

[3] TNO, partner of Solliance, High Tech Campus 21, Eindhoven 5656 AE, the Netherlands

[4] Molecular Materials and Nanosystems, Institute for Complex Molecular Systems, Eindhoven University of Technology, partner of Solliance, P.O. Box 513, 5600 MB Eindhoven, the Netherlands

**Abstract**

A spectral splitting, light trapping dielectric metasurface is designed, fabricated and integrated into a four-terminal perovskite/silicon hybrid tandem solar cell to increase the absorption of light close to the bandgap of the perovskite top cell, and enhance transmission of the near-infrared spectral band towards the bottom cell. The metagrating is composed of a hexagonal array of unit cells of 150-nm-tall hydrogenated amorphous silicon trimer nanostructures with dielectric Mie resonances in the 600-800 nm perovskite near-gap region, made using substrate-conformal imprint lithography. By tailoring the metasurface resonant scattering modes and their interference with the direct reflection paths we minimize specular reflection and obtain high diffraction efficiency that leads to improved light trapping in the perovskite top cell. The measured short-circuit current increase in the perovskite top cell is 0.5 mA/cm² corresponding to an estimated efficiency gain of 0.26% (absolute) for the metasurface-integrated 4T perovskite/silicon tandem cell. Simulations for a further optimized metasurface spectrum splitter geometry predict a short-circuit current gain in the perovskite top cell of 1.4 mA/cm² and an efficiency gain for the 4T tandem cell of 0.4% (absolute). The metagrating approach for simultaneous spectral splitting, light trapping and reflectance reduction provides a flexible platform that can be applied to many tandem cell geometries.

**Introduction**

Perovskite/silicon tandem cells are promising candidates for high efficiency solar cells that could find their route to broader commercialization in the next years. The present efficiency record of 29.15% lies well above the single-junction Si solar cell efficiency and the record will likely increase further in the coming years [1]. Four-terminal (4T) tandem cells are of specific interest, as for this cell design current matching is not needed and therefore their efficiency is less dependent on the top cell bandgap than for the case of two-terminal (2T) cells. 4T tandem cells are composed of two independent cells, with a transparent bottom contact on the top cell to allow light transmission to the underlying cell. In the top cell light absorption is usually incomplete, as the cell thickness is typically on the order of 0.5 µm, and part of the light close to the bandgap of the top material is transmitted to the underlying bottom cell. A schematic of the absorption of the solar spectrum in the top and bottom cell of a 4T perovskite/silicon tandem cell is shown in Figure 1a, using measured data for the transmission from a perovskite top cell [2] and full absorption in the silicon bottom cell. Light with wavelength in the λ=600-800 nm spectral range at energy just above the perovskite bandgap energy (1.55 eV in this case) is partly transmitted and then absorbed in the underlying silicon cell. This creates losses in the tandem cell, as light close to the bandgap absorbed in the bottom cell creates a lower photovoltage than in the top cell. Introducing a spectral splitter interlayer in between the top and bottom cells can help circumvent these losses [3]. Ideally, a spectral splitter is designed such that all light with energy above the bandgap that is not absorbed in the first pass through the top cell is reflected back and trapped in the top cell. At the same time the spectral splitter can help reduce reflection of infrared (IR) light at the top/bottom cell interface, and thereby create enhanced IR absorption in the underlying silicon cell [4]. In our 4T tandem cell design [2] the perovskite top cell is placed on top of the silicon bottom cell with an airgap in between. This geometry allows to put the spectral splitter placed on a glass slide on the bottom of the perovskite cell without modification of the silicon bottom cell as the schematic in Figure 1b shows. To not introduce additional losses in the silicon cell, the spectral splitter/air interface has to fulfill the requirement to have a lower near-infrared reflectance than the glass/air interface on the perovskite bottom side in the top cell.

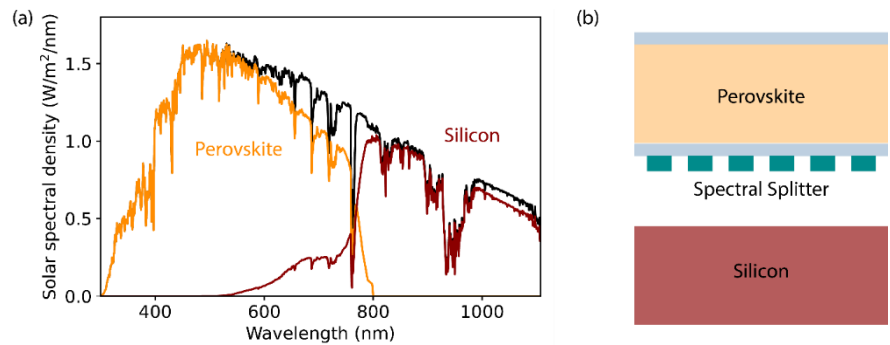

**Figure 1** 4T Perovskite/silicon tandem solar cell **(a)** Absorption of solar spectrum (black) in 500 nm thick perovskite top cell (orange) and silicon bottom cell (red) based on experimental data from [2]. **(b)** Schematic of air-coupled 4T tandem cell with spectral splitter metasurface on bottom side of perovskite top cell. Blue bands on top and bottom of the perovskite are glass layers (not to scale); the top glass is the superstrate for perovskite growth; the bottom glass layer supports the metasurface and is attached to the perovskite with index-matching oil.

Spectral splitting has been explored in different tandem cell concepts and other hybrid solar cell systems. The spectral splitter was often designed to split the spectrum and direct the light in a specular direction. Most common approaches are the use of (semi-)reflective elements such as multilayer structures [5] and dichroic mirrors[6]–[9], or selective reflection off the top cell [10]. Intermediate reflectors for improved absorption in the top cell have also been studied before, using Bragg reflectors [4], [11]–[14] which can control the splitting spectrum in a stacked tandem cell and optimize for reduced IR reflection at interfaces. However none of these planar designs can direct light over an angular range to create light trapping. Martins *et al.* studied intermediate structures in 4T tandem cells and concluded that optical impedance matching between the sub cells is of higher importance than spectral splitting, however, not taking light trapping into account [4]. Three-dimensional photonic crystal structures have been used as intermediate reflectors with a small light trapping effect introduced by their diffraction orders [15], [16], but additional anti-reflection coatings are necessary to keep these structures effective [17].

In this work, we design a spectral splitter that acts as a semi-transparent light trapping layer for the perovskite top cell. It reflects light with energy above the perovskite bandgap back to the top cell under high diffraction angles, creating light trapping in the top cell, while the long-wavelength spectral range is transmitted nearly lossless to the bottom cell. Previously, it has been shown that for a spectral splitter in a 4T tandem to be effective light has to be reflected in a Lambertian fashion to achieve maximum light trapping in the top cell [3]. However, the challenge has remained of how to create the spectral selectivity of Lambertian scattering that is required for the 4T design. For example, conventional textured scattering surfaces can create Lambertian light trapping but do not have spectral control.

Here, we introduce wide-angle scattering light-trapping metasurface that is spectrally selective by making use of dielectric scattering nanoparticles with optical Mie resonances. Dielectric metasurfaces have been used previously to improve the performance of tandem solar cells for increased absorption and anti-reflection from the different interfaces [18], [19], and light trapping [20], [21]. First, we recap the theoretical 4T tandem efficiency limits in the thermodynamic limit case, and investigate potential improvements using an optimized metagrating. We then design and fabricate the metagrating and integrate it into a 4T perovskite/silicon tandem device [2]. In comparison with a 4T reference without spectrum splitter we measure a current improvement of 0.5 mA/cm² in the top cell of the tandem. The metasurface spectral splitter offers the first experimental solution to a spectral splitting layer with high light trapping in 4T tandem cells.

**Thermodynamic limit for spectrum splitter designs**

In this work, we distinct between a planar and a Lambertian spectral splitter (see schematics in Figure 2a). The planar spectral splitter reflects light back specularly and creates a single extra path of absorption through the top cell. The Lambertian spectral splitter reflects light back following a Lambertian angular distribution and creates enhanced light trapping in the top cell. We first calculate the thermodynamic detailed-balance efficiency limits, assuming imperfect absorption in a 580-nm-thick perovskite top cell and unity transmission to the underlying silicon cell using the methods discussed in [22]. Auger recombination in Si and other non-ideal processes are not taken into account and we assume full carrier collection in top and bottom cells. While the maximum efficiencies calculated using the detailed-balance model cannot be reached in reality, they allow a reliable comparison between different configurations and help predict important trends.

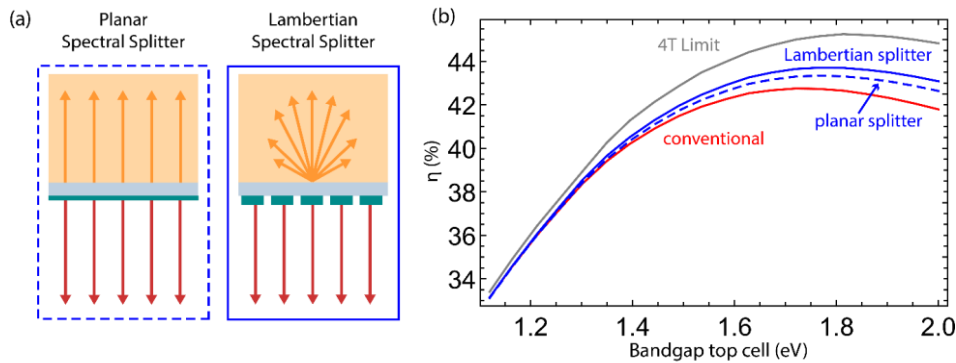

**Figure 2** Spectral splitting in 4T tandem cell. (a) Schematics of planar (left) and Lambertian (right) spectral splitter. (b) Efficiency limit from detailed balance calculations of 4T tandem cell as function of top cell bandgap energy with silicon bottom cell ($E_{BG}$=1.12 eV). Thermodynamic limit (gray line), limit for 580 nm thick perovskite top cell (red line) and 580 nm thick perovskite top cell with integrated planar (dashed blue line) and Lambertian spectral splitter (solid blue line).

The results for the planar 580-nm-thick top cell, as well as the integrated planar and Lambertian spectral splitter are plotted in Figure 2b. For both configurations we assume ideal spectral splitting, with 100% reflection of light with energy above the bandgap of perovskite (see details in supplementary) and unity transmission into the bottom cell below that range. As seen in Fig. 1, the earlier conventional tandem design suffered from significant reflection losses at the perovskite/Si interface, as well as parasitic absorption in the transparent contacts, resulting in incomplete light collection in the Si bottom cell.

As can be seen in the figure, for a top cell bandgap of $E_{BG}$=1.55 eV (used in our previous work [2]), an absolute increase in efficiency of 0.3/0.6 % is found for the planar/Lambertian spectral splitter, respectively. Enhancements by more than 1% are found for larger perovskite bandgaps. These calculations for the thermodynamic limiting case pave the way for designs of realistic geometries that are discussed next.

**Spectrum splitting metasurface design**

In this work scattering nanoparticles are arranged in a metagrating structure, a geometry that scatters light only in a well-defined resonant scattering spectral band, towards an optimized distribution of scattering angles, as we have shown before [23]–[25]. We design the scattering resonance spectrum of trimer-assemblies of resonant dielectric Mie scatterers that form the unit cell of the grating. The angular directivity is controlled by the periodicity of the grating in combination with the angular scattering distribution of the unit cell. We optimize the metagrating such that specular reflection is cancelled out, by creating destructive interference between light scattering from the resonant Mie particles and light reflected from the glass substrate that holds the scatterers. As scattering nanoparticles we use hydrogenated amorphous silicon (a-Si:H) cylinders; their high refractive index (n=3.89 at $\lambda$=750 nm) makes them efficient scatterers, while low losses in the near-infrared (k<0.01 at $\lambda$=750 nm) make them transparent in the perovskite light trapping range. The particle geometries are arranged on a 0.7 mm thick glass slide in a hexagonal grating.

To integrate the metasurface with the perovskite top cell, the glass slide is placed with index matching liquid against the indium tin oxide (ITO) bottom contact of the cell, with the nanopattern facing the air gap towards the silicon bottom cell. The perovskite top cell is grown in a superstrate configuration, with a 0.7 mm glass plate on its front side (see Fig. 1). The metasurface is designed and positioned in such a way that light trapping occurs in between the top and bottom glass plates of the top cell. Light that is scattered within the escape cone leaves the front of the top cell after one extra path through the perovskite. Light scattered from the metagrating outside the escape cone is reflected from the top glass back into perovskite. When interacting with the metasurface again, it partly leaves the cell on the bottom side, is partly scattered inside the cell again and, by reciprocity, partly scattered upwards, leaving the top cell on the front side.

The design of an optimized spectral splitter requires an integrated optimization of several scattering mechanisms. First, the Mie scattering unit cells must be made such that the resonance spectrum matches the desired $\lambda$=600-800 nm light trapping spectral band for the perovskite. It must also be completely off-resonant for longer wavelength to facilitate full transmission of the 800-1200 nm spectral band into the silicon bottom cell. Furthermore, the scattering cross section combined with the geometrical unit cell fill fraction must create near-unity interaction of incident light with the metagrating in the 600-800 nm spectral band. We use insights from our previous work on metagratings [23], [24] to design the far-field interference of the electric and magnetic dipole and quadrupole Mie resonances to create the desired spectral shape and scattering strength. In parallel, we tailor the intensity and phase of the specularly scattered light such that it destructively interferes with light reflected off the glass/interface.

Our metasurface design is based on a hexagonal grid to have diffraction into 6 different azimuthal directions. The periodicity p=525 nm was chosen to create diffraction in glass below $\lambda$=800 nm. Light with a wavelength larger than $\lambda$=525 nm has a diffraction angle larger than 42° inside glass and therefore experiences internal light trapping in the top cell. We perform finite-difference time-domain (FDTD) simulations [26] to calculate the reflection/transmission spectra of the metasurface on glass numerically. In the simulations, the source was placed inside the glass substrate, together with a reflection monitor behind it, in the glass, to determine the angle and intensity of individual diffracted orders inside the glass-perovskite-glass stack, while the transmission monitor was placed outside the bottom glass carrying the metasurface, to calculate the transmission to the underlying silicon cell. The dimensions (height and diameter) and unit-cell geometry (arrangement of scattering particles) were optimized for maximum efficiency of the complete 4T tandem cell. The optimized geometry consists of a unit cell composed of three 170-nm-tall scattering particles arranged in a windmill-like trimer geometry with elliptical wings of 70 nm diameter and 140 nm length (see SI for details on the geometry).

The simulated reflection spectrum of the optimized metasurface is plotted in Figure 3a. The total reflectance is shown as well as the part that is diffracted into diffraction orders. One can see that the a-Si:H nanoparticles reach a resonant reflectivity up to 80% in the $\lambda$=600-800 nm wavelength range, with a peak at around 750 nm, where enhanced light coupling and trapping in the top cell is desired. The major part of the reflected light is diffracted and trapped in the perovskite top cell. Furthermore, the reflectance for wavelengths between 800–1200 nm is around 0.5-2.5%, which is well below the reflectance of the glass-air interface (4%). This reduced reflectance for higher wavelengths is due to the fact that the particles are designed to show enhanced forward scattering in this wavelength regime, in addition to the fact that the metasurface has an effective index of around n=1.1 between $\lambda$=800–1200 nm and therefore acts as an effective anti-reflection coating for glass [27]. The absorption spectrum is also shown in Fig. 3a and shows a Mie-resonant absorption peak for the a-Si:H nanoparticles as well as an increasing absorption due to tail states in the a-Si:H electronic bandstructure. This absorption forms no limitation in the spectrum splitting operation as the perovskite top cell is absorbing well in the low-wavelength range.

To determine the potential overall gain in efficiency due to integration of the spectral splitter we first use the detailed-balance calculations with perovskite bandgap 1.55 eV to determine the change in absorption in the top and bottom cell for the geometry in Fig. 1b. We compare a perovskite top cell with a glass bottom slide with the same cell with the above described spectral splitter metasurface. We take into account absorption enhancement in a 580-nm-thick perovskite top cell due to spectral splitting and light trapping, the corresponding absorption reduction in the silicon bottom cell and the absorption gain in the bottom cell due to higher transmission in the IR. The efficiency that can be gained is 0.7% absolute compared to the same tandem cell with a glass-air interface without spectral splitter. Interestingly, this is higher than the 0.6% increase calculated above for the idealized Lambertian spectrum splitter. This is because the reduced IR reflectance to the planar glass-air interface is also taken into account here, while it was not in the calculations above. The short-circuit current gain in the perovskite top cell is calculated to be 1.6 mA/cm², a relative increase of 7%. Translating this to the experimental short circuit current for the perovskite top cell that we will use below [2] this corresponds to a short-circuit current enhancement of 1.4 mA/cm$^2$.

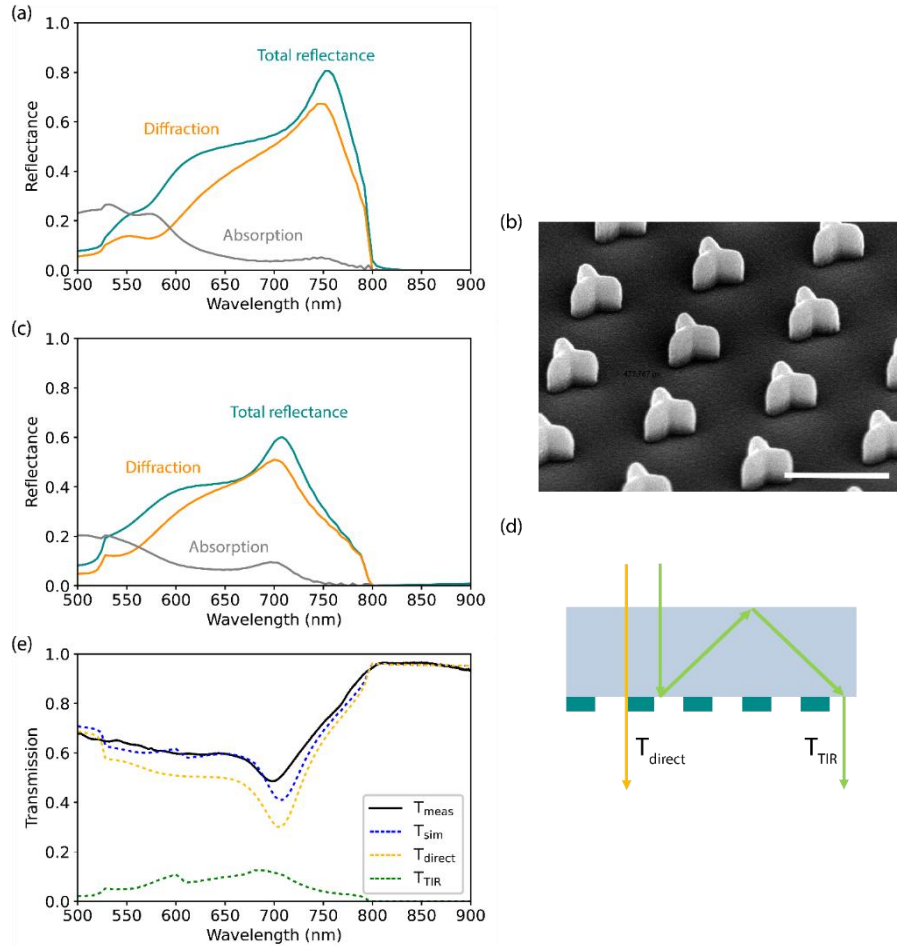

**Figure 3** Spectral splitter metasurface. (a) Simulated total reflectance (green), diffraction (orange) and absorption (grey) of optimized spectral splitter metasurface. (b) SEM tilted top-view image of metasurface: a-Si:H windmill shaped particles on glass substrate (covered with around 10 nm chromium layer for better imaging). Scale bar is 500 nm. (c) Simulated total reflectance (green), diffraction (orange) and absorption (grey) of fabricated spectral splitter metasurface. (d) Paths of transmission through glass slide with same coloration as in (e). (e) Optical transmission measurements (black) of fabricated metasurface, calculated/simulated transmission (blue dashed line) composed of direct transmission (yellow dashed line) and transmission after total internal reflection (green dashed line).

**Metagrating spectrum splitter fabrication**

To fabricate the spectrum splitting metasurface and integrate it in the 4T design, a 150-nm-thick a-Si:H layer was deposited on a glass substrate by plasma-enhanced chemical vapor deposition (PEVCVD), and the glass was cut into 2×2 cm² samples. We use Substrate Conformal Soft Imprint Lithography (SCIL), to replicate the structure over an area of several cm² [28]. For this, first we create a SCIL stamp with an imprint area of 2.5×2.5 cm² with notches with parameters of the nanoparticles described above. We spin-coat a layer of 50-nm-thick silica sol-gel on top of the a-Si:H/glass sample, imprint the sol-gel with the SCIL stamp and let it cure for 8 minutes. After removal of the stamp, the pattern is transferred into the silicon with a double Reactive Ion etching (RIE) plasma etch step, first to break through the sol-gel layer (25 sccm $CHF_3$, 25 sccm Ar) and second to transfer the pattern from the sol-gel mask into the silicon layer (15 sccm $CHF_3$, 10 sccm $SF_6$, 3 sccm $O_2$). The residual of around 10 nm of sol-gel on top of the Si nanoparticles showed a negligible effect on the results in simulation and was therefore not removed from the silicon pillars. Figure 3b shows a Scanning Electron Microscopy (SEM) top-view image of the fabricated structure. It matches quite well with the targeted dimensions calculated above: the windmill wings are 15 nm shorter than designed and the height of the particle is 150 nm, 20 nm less than designed. Figure 3c shows the simulated reflectance of the fabricated structure. The reflectance shows a maximum above 60% and the resonance peaks near 700 nm, somewhat blue-shifted from the desired range. However, the simulation shows also for this fabricated geometry the key desired features: high diffraction efficiencies and low reflectance above 800 nm. Using the simulated data for the experimental geometry as input for detailed-balance calculations we calculate an efficiency gain for the perovskite/Si tandem with metagrating in Fig. 3b of 0.6% absolute, and a current gain of 1.0 mA/cm² (+4%) in the perovskite top cell. In practice, the latter would translate to a 0.9 mA/cm$^2$ enhancement in the experimental perovskite top cell.

**Experiments**

To characterize the fabricated spectrum splitter, two different measurements were performed. First, the transmission of the metasurface/glass sample was measured and compared with the simulations. Second, the spectral splitter was integrated with a semi-transparent perovskite top cell using index-matching fluid and External Quantum Efficiency (EQE) as well as transmission and reflection measurements were performed. In parallel, we used measurements of the external quantum efficiency of the Si bottom cell to calculate the performance of the metasurface-integrated 4T tandem cell geometry.

**Optical measurements.** Transmission measurements of the fabricated metasurface/glass sample were performed in an integrating sphere setup with an NKT super-K white-light laser source with a collimated beam at perpendicular incidence on the sample. The measured transmission is mainly composed of two contributions (Figure 3d); direct transmission though the glass slide with metasurface, and transmission of light that follows diffracted paths inside the glass. The measured transmission is shown in Figure 3e and compared with the sum of the contributions derived from the simulated spectrum of the fabricated sample. The spectrum compares very well with the simulated transmission spectrum, in which we have taken into account an additional 4% reflection at the top glass surface and total internal reflectance inside the glass slide. The measured transmission dip of 35% near ~700 nm in Fig. 3e indicates the metasurface effectively reflects in the spectral range near the perovskite bandgap, in agreement with the simulated reflectivity in Fig. 3c. The simulations show that up to ~12% of the transmitted light originates from paths that are first diffracted off the spectrum splitting metasurface. For a metasurface integrated with the perovskite top cell these paths will lead to enhanced absorption in the perovskite. Given the fact that strong correspondence is achieved between simulation and experiment in Fig. 3e, we are confident that optimized metasurface design presented above (with more redshifted resonance and higher resonant reflectivity) will in practice yield the 1.4 mA/cm$^2$ current enhancement for the perovskite top cell as calculated above.

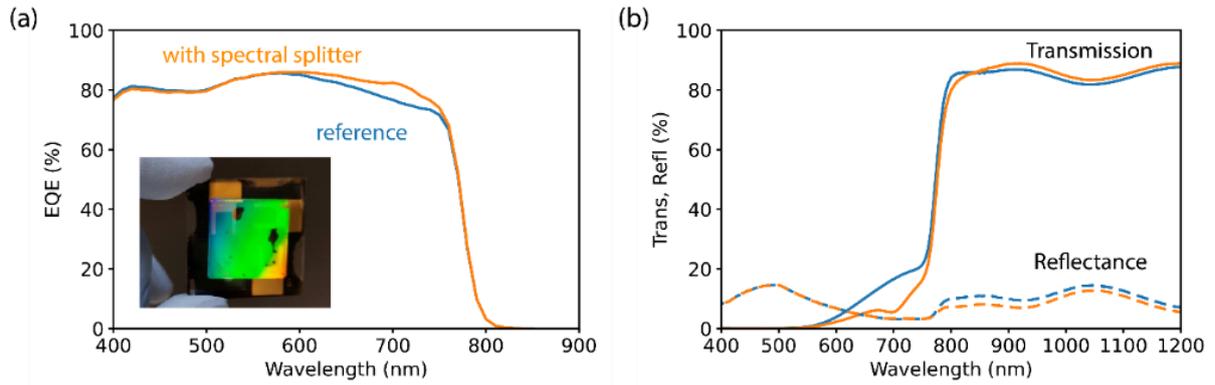

**Figure 4** Measurements of metasurface integrated with perovskite cell. (a) EQE measurements of perovskite with spectral splitter (orange) and with glass reference (blue) Inset: Photograph of perovskite with spectral splitter taken from the bottom (facing the metasurface). Diffracted colors from the metasurface are clearly visible. (b) Transmission and reflectance measurement of the perovskite cell with spectrum splitter integrated.

**Transmission and reflection spectra and EQE measurements of spectral splitter integrated in perovskite cell.** Next, the glass/metagrating sample was placed on the bottom side of a perovskite top cell (see schematic of Figure 1b, and inset in Figure 4a). The glass side of the sample was connected with index matching oil on the ITO bottom side, with the silicon metasurface facing towards air. The perovskite cell is composed of a ~580 nm thick $Cs_{0.05}(MA_{0.17}FA_{0.83})Pb(I_{0.9}Br_{0.1})_3$ absorber layer with hole and electron transport layers and ITO top and bottom contact layers. The layer stack is composed (from top to bottom) of thermally-evaporated $MgF_2$, a Corning XG glass substrate, sputtered hydrogen doped indium oxide, spincoated PTAA, the spincoated perovskite layer, evaporated C60, a layer of $SnO_2$ deposited with spatial atomic layer deposition, and another layer of sputtered ITO.[2] For the reference measurements a perovskite cell with a glass bottom slide was used. EQE measurements were are shown in Figure 4a. The metasurface spectrum splitter creates a clearly enhanced EQE over the 600-750 nm spectral range, corresponding to a current gain of 0.4 mA/cm² in the perovskite top cell. This trend is in agreement with the reduced transmittance of the cell in this spectral range as shown in Figure 4b. The transmittance spectrum shows an enhanced transmission above λ=800 nm demonstrating the effective forward scattering and anti-reflection effect of the metagrating. No additional reflectance is measured below λ=800 nm showing the spectrum splitter does not create additional escape of light from the top side of the perovskite. Above λ=800 nm, the reflectance is lower than in the reference sample, consistent with the increased transmission.

Finally, we compare the gain in short-circuit current derived from Fig. 4a (0.4 mA/cm²) with the simulations. As described above, the detailed-balance calculations, scaled to the experimental short-circuit current for the used perovskite predict a current improvement of 0.9 mA/cm², well above the measured value. We attribute the difference between measurement and simulation to light leaking from the edge of the spectral splitter, as well as incomplete light trapping because light can escape from the edges of the top cover glass. In the present geometry the perovskite cells size is only 4x4 mm². As light close to λ=800 nm is refracted from the metagrating under a wide angle a major fraction of it will escape from the 0.7-mm-thick glass slide and is thus not absorbed in the perovskite cell. Furthermore, light that passes through the perovskite cell once and experiences total internal reflection on the top glass slide is not channelled through perovskite a second time, because of the small cell size (see sketch and further explanation in SI). To qualitatively prove the importance of this effect we repeated the EQE measurement with a similar sample that was cut to roughly match the size of the perovskite top cell and covered the sample edges with a ~50-nm-thick silver layer to serve as a mirror. With this modification we measure (from EQE measurements) a slightly higher current increase of 0.5 mA/cm² in the perovskite top cell. Given the strong correspondence between reflectance simulations and measurements (Fig. 3b) further improved metasurface/cell integration is expected to yield the estimated current gain of 0.9 mA/cm². Taking into account the EQE of the used Si bottom cell this would translate into an overall efficiency gain of the 4T tandem cell with spectral splitter of 0.26% (abs.). (See SI). Calculations using the further optimized metasurface design introduced above yield an efficiency enhancement for that geometry of 0.40%. At last, it should be mentioned that adding an anti-reflection coating of $MgF_2$ as in [2] on the bottom side of perovskite can lead to further enhanced transmission in the infrared compared to the spectral splitter geometry in this work. Such additional optimization steps for reduced reflectance in the infrared are therefore of interest in future.

**Conclusion**

In conclusion, we designed and fabricated a spectral splitting light trapping dielectric metasurface as an interlayer in 4T perovskite/silicon tandem cells. We use the resonant light scattering principle of metagratings to design a spectral splitter with high diffraction efficiency in the bandwidth of interest and reduced reflectance and enhanced transmittance for the near-infrared spectral band. Simulations for the experimental geometry, corrected for experimental cell parameters, predict a short-current current enhancement in the perovskite top cell of 0.9 mA/cm² and an efficiency gain of 0.26% (abs.). The experimental enhancement derived from EQE measurements is 0.5 mA/cm², the difference partly attributed to light leakage due to the small sample size. Simulations show a further improved design could yield an experimental short-circuit current enhancement of 1.4 mA/cm², corresponding to an efficiency enhancement of 0.40% (abs.). The demonstrated metagrating structure is flexible in design and can be adjusted to other top bandgaps or tandem geometries. The soft imprint technique to fabricate the metagratings is scalable up to wafer-sized imprints. Our work shows that the critical balance between spectral splitting, light trapping and concurrent reduced reflectance in the infrared can be fulfilled with these dielectric metasurfaces.

**Acknowledgements**

# Supplementary Information

# Four-terminal perovskite/silicon tandem solar cell with integrated Mie-resonant spectral splitter metagrating

Verena Neder[1,2], Dong Zhang[3,4], Sjoerd Veenstra[3], and Albert Polman[2]

[1]Institute of Physics, University of Amsterdam
Science Park 904, 1098 XH Amsterdam, the Netherlands

[2]Center for Nanophotonics, NWO-Institute AMOLF
Science Park 104, 1098 XG, Amsterdam, the Netherlands

[3] TNO, partner of Solliance, High Tech Campus 21, Eindhoven 5656 AE, the Netherlands

[4] Molecular Materials and Nanosystems, Institute for Complex Molecular Systems, Eindhoven University of Technology, partner of Solliance, P.O. Box 513, 5600 MB Eindhoven, the Netherlands

**1. Ideal design conditions of spectral splitter for an ideal 4T tandem solar cell with Si bottom cell**

We define the reflectivity $R(E)$ of an ideal spectral splitter as a step function with tunable step $\Delta E$ away from the bandgap energy of the top cell and reflectance $r$:

$$R(E) = \begin{cases} 0, E < E_{BG} + \Delta E \\ r, E \geq E_{BG} + \Delta E \end{cases} \quad (S1)$$

In Figure S1 the parameters $r$ and $\Delta E$ are swept to find the optimal splitting conditions, leading to the maximal efficiency enhancement compared to a 4T tandem cell with Si bottom cell without spectral splitter, as shown in the main text in Figure 2b. The top cell bandgap was fixed to $E_{BG}$=1.55 eV. The ideal parameters were found to be $r$=1 and $\Delta E$ =0.1 eV and $\Delta E$ =0.06 eV for the planar and Lambertian spectral splitter respectively. The smaller value for $\Delta E$ found for the Lambertian case reflects the better light trapping for that geometry.

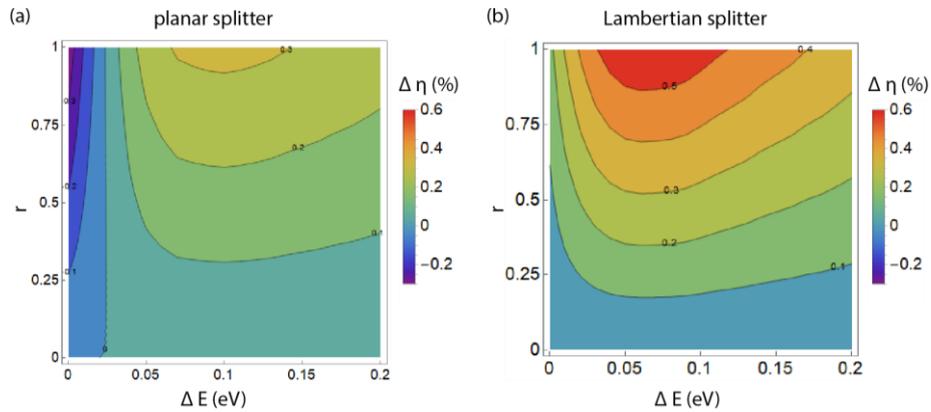

Figure S2 Parameter sweep of selective reflector step function for a (a) planar and (b) Lambertian spectral splitter in a 4T tandem solar cell with a top cell bandgap of E=1.55eV and Si bottom cell.

## 2. Optimized spectral splitter design

The unit cell of the metasurface was optimized using parameter sweep optimization. The windmill-shaped particle described in the main text is composed of 3 identical ellipses (see Figure S2a), each rotated by 120°, with minor radius of $r_1$ and major radius $r_2$. The relation between $r_1$ and $r_2$ is given by $r_2 = f_1 \cdot r_1$. The distance between the center of the ellipses ($C_1$, $C_2$ and $C_3$) and the center of the unit cell ($C_0$) is given by d, with d given by : $d = r_1 + f_2 \cdot r_1$. The maximum gain in efficiency for the metasurface-integrated 4T geometry was found for a hexagonal lattice with pitch 525 nm and a unit cell described by : $r_1$=35 nm, $f_1$=2.4, and $f_2$=0.7. The simulated scattering cross section of this windmill-shaped particle normalized by the geometrical area is shown in Figure S2b. The particle has a clear resonance in the relevant wavelength band between 550-800 nm with a peak scattering cross section beyond 10. This implies that a geometrical filling fraction of 10% suffices to achieve full interaction of incident light with the metasurface on resonance.

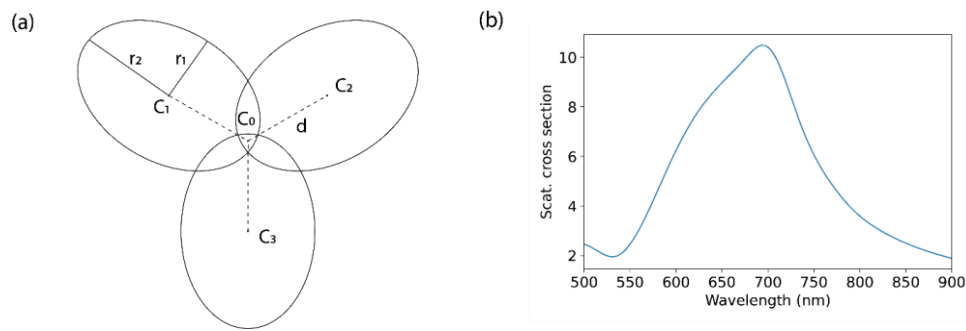

Figure 3 (a) Dimensions of the windmill shaped a-Si:H particle in the hexagonal unit cell of the metagrating. (b) Scattering cross section of the optimized particle.

## 3. Edge effect and overall tandem cell performance with integrated spectral splitter metagrating

The spectral splitter metasurface was designed for a wafer-sized solar cell where edge effects do not play a major role. The simulations were performed not taking any edge effects into account. However, in experiments, the size of the perovskite top cells is limited to 4 mm x 4 mm. This means that, for the glass thickness that is used (0.7 mm), light with wavelengths above 760 nm, which has a diffraction angle below 75° is lost at the edges. This is a substantial part of the reflectance spectrum and can explain the lower measured current gain in the perovskite, in comparison to the simulations (see Fig. S3 top schematic, red solid light path). Additionally, light that is reflected to the perovskite top cell and experiences total internal reflection on the top glass surface also is partially lost due to the small size of the perovskite and does not experience a second pass through the cell (see Fig. S3 top schematic, red dashed light path). The geometry with the Ag coated edges described in the main text is presented in the bottom schematic of Figure S3 and can partly reduce the edge effect. Furthermore, in the EQE measurements the light beam has a 3 mm–diameter spot size, so that part of the light that is diffracted under shallower angles can escape from the edges, lowering the achievable current enhancement.

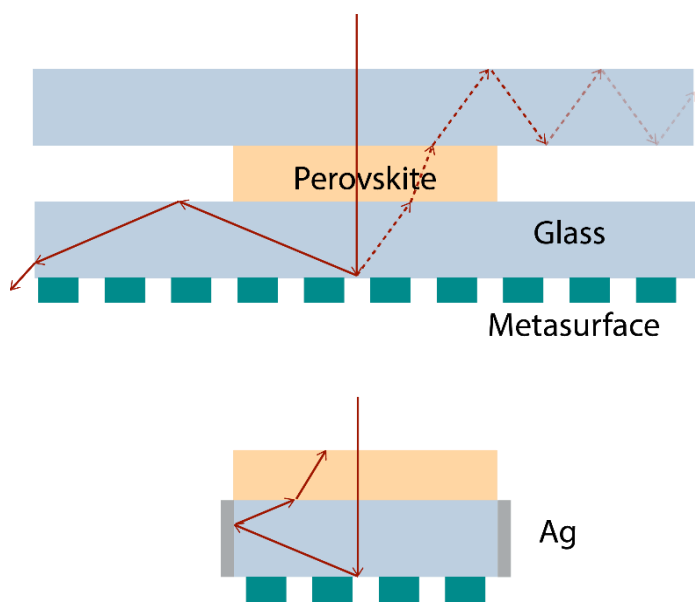

Figure S4 Schematic of edge effect in experiment. Top schematic: Part of the light escapes because the spectral splitter is bigger than perovskite cell (red solid line) and because the front glass slide is bigger than the perovskite cell (dashed red line). Bottom: If the spectral splitter is cut and the edges are covered with silver light is redirected to the perovskite cell

**3. Overall tandem cell performance with integrated spectral splitter metagrating**

Figure S4 shows EQE measurements of the silicon solar cell in the 4T geometry with the spectrum splitter integrated in the top cell. We use a commercial 5" interdigitated-back-contact (IBC) c-Si solar cell from SunPower. The data are compared to the EQE of a reference 4T tandem cell with a glass plate on the bottom side of the perovskite. For wavelengths below 800 nm the EQE in the silicon cells is decreased, because, as designed, light is reflected towards the perovskite top cell by the spectral splitter. Above 800 nm, the EQE is increased compared to the reference sample, due to reduced near-IR reflectance at the bottom side of the perovskite cell by the spectral splitter.

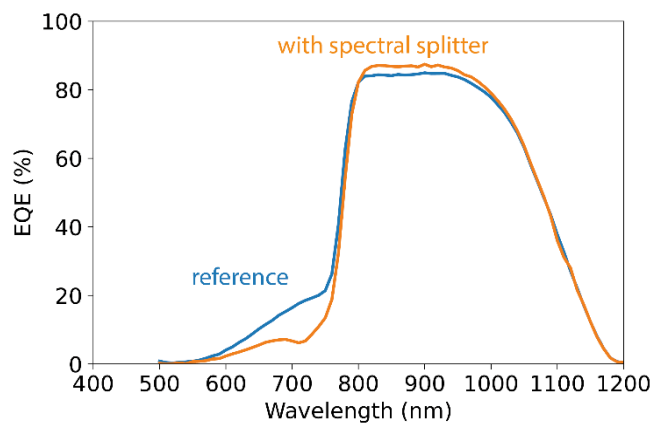

Figure S5 EQE measurement of Si bottom cell with and without spectral splitter.

From the EQE data we infer that a short-circuit current of 0.8 mA/cm$^2$ is lost in the silicon cell. With the presently measured 0.5 mA/cm$^2$ short-circuit current gain in the perovskite top cell this would not enhance the tandem cell efficiency. However, as described above light leakage in the small sample geometry has likely affected the current measurement. Assuming the calculated 0.9 mA/cm$^2$ current enhancement for the experimental geometry (see main text), an efficiency gain for the tandem due to the spectrum splitter of 0.3% is expected. As described in the main text, a further optimized metasurface geometry would lead to a 1.4 mA/cm$^2$ short-circuit current enhancement in the perovskite cell, which would translate into an estimated 0.4% efficiency enhancement for the tandem cell. For these efficiency calculations, the open-circuit voltage ($V_{oc}$) and fill factor (FF) were taken from I-V measurements of the individual cells in the 4T geometry, as listed in Table S1. The Table also summarizes the efficiencies for different geometries quoted in the paper.

Table 1 Performance of the perovskite, c-Si cell and tandem cell with and without spectral splitter. The enhancements due to the spectrum splitting metasurface compared to the geometry without spectrum splitter are indicated between brackets.

| Cell | | $V_{oc}$ (mV) | $J_{sc}$ (mA/cm$^2$) | FF | Efficiency (%) |
|---|---|---|---|---|---|
| **perovskite top cell** | reference | 1063 | 20.7 | 0.778 | 17.12 |
| | with spectrum splitter + Ag | 1063 | 21.2 (+0.5) | 0.778 | 17.53 |
| | | | | | |
| **Si bottom cell** | reference | 703 | 16.1 | 0.800 | 9.05 |
| | with spectrum splitter + Ag | 701 | 15.3 (-0.8) | 0.799 | 8.57 |
| | | | | | |
| **4T perovskite/Si tandem: experiment** | reference | | | | 26.17 |
| | with spectrum splitter | | | | 26.10 (-0.07) |
| | | | | | |
| **4T perovskite/Si tandem: calculations** | reference | | 20.7 (perovskite) | | 26.17 (4T cell) |
| | with fabricated splitter | | 21.6 (+0.9) (perovskite) | | 26.43 (+0.26) (4T cell) |
| | with optimized splitter | | 22.1 (+1.4) (perovskite) | | 26.57 (+0.40) (4T cell) |